\documentclass[10pt]{article}
\usepackage{multicol,citesort,epsfig,overcite}
\begin{document}

\title{Molecular dynamics of a dense fluid of polydisperse hard spheres}

\author{{\bf Richard P. Sear}
~\\
Department of Physics, University of Surrey\\
Guildford, Surrey GU2 7XH, United Kingdom\\
email: r.sear@surrey.ac.uk}

\maketitle

\begin{abstract}
Slow dynamics in a fluid are studied in one of the most basic systems
possible: polydisperse hard spheres. Monodisperse hard spheres
cannot be studied as the slow down in dynamics as the density
is increased is preempted by crystallisation.
As the dynamics slow they become more heterogeneous, the spread
in the distances traveled by different particles in the
same time increases. However,
the dynamics appears to be less heterogeneous than in hard-sphere-like
colloids at the same volume fraction.
The particles which move least far
in a characteristic relaxation time
and, particularly, the particles which move farthest in
the same time are clustered, not randomly distributed throughout the sample.
We study the dynamics at three different widths of the distribution
of diameters of the hard spheres. For each width, the
relaxation time is the same function of the compressibility factor,
suggesting that this determines
the relaxation time for hard spheres.
\end{abstract}

%\vspace{0.2in}
%PACS: 82.70.Dd, 81.30.-t, 64.70.Dv

%82.70.Dd Colloids
%64.70.Dv Solid-liquid transitions
%81.30.-t Phase diagrams and microstructures developed by solidification 
%and solid-solid phase transformations

%\newpage
\begin{multicols}{2}

\section{Introduction}

The study of hard spheres has taught us much about the equilibrium
and dynamic behaviour of dense fluids and crystals. We have learnt
that crystallisation can be a result of the more efficient
packing in the crystal of atoms or colloids,
and that diffusion is inhibited in dense fluids by repeated collisions
of particles with their neighbours. One outstanding problem is that
of understanding and perhaps trying to predict, the dramatic
slowdown of the dynamics of fluids and the formation of a glass; 
see Refs. \citen{ediger96,angell95,debenedetti}
for introductions to slow dynamics and glasses.
An obvious thing to do is to study this phenomenon in hard spheres.
This has been done and the results are presented here.

Unfortunately, monodisperse hard spheres crystallise readily,
preventing the study of hard spheres at densities high enough to
observe very slow dynamics. Crystallisation is inhibited
if the spheres are not all the same size. Speedy studied a binary mixture of
hard spheres of two diameters \cite{speedy98}, which did
not crystallise.
Here mixtures
of spheres with continuous ranges of diameters are studied, these are called
polydisperse mixtures.
This will enable us to compare with experiments
on hard-sphere colloids, which are inevitably on polydisperse
particles, \cite{kasper99,kegel00,weeks00}.
Monte Carlo simulations of polydisperse
hard spheres have been performed by Doliwa and Heuer
\cite{doliwa98,doliwa99}.

In this contribution we will quantify the slow down in the dynamics,
characterise these slow dynamics, and assess the effect of having
spheres of different diameters present.
The slow down is measured by calculating a relaxation time.
The dynamics are characterised by calculating their deviation from
what we would expect if each particle was diffusing independently.
When the deviations are large we call the dynamics heterogeneous.
When the dynamics are heterogeneous, the distribution of speeds
and relaxation times of individual particles is much broader
than the Gaussian function obtained when particles are
diffusing independently. There are many more particles
which have traveled much farther than the mean than
we would expect for independent diffusion of each particle.
These fast-moving particles are not distributed randomly in space, they
are concentrated in clusters. In other words,
fast particles tend to be
surrounded by other fast particles
\cite{hurley95,kob97,donati99b,glotzer99}.

Three different widths of the distribution of diameters are studied.
The polydispersity width is characterised by the
standard deviation of the distribution in diameters.
For the three distributions this is 9, 14 and 20\%. As in
experiments on hard-sphere colloids \cite{pusey91,megen98}
if the distribution is narrower the spheres can crystallise.
Surprisingly, although increasing the polydispersity at fixed
volume fraction speeds up relaxation, the relaxation time
is, within simulation resolution, independent of the width of
the distribution at fixed compressibility factor. The relaxation time
appears to depend only on the compressibility factor. We also study the
effect of the heterogeneity in diameters on the heterogeneity
in the dynamics.

In the next section we briefly describe our, standard, methodology.
Section 3 outlines when our simulations
crystallised and quantifies how much polydispersity in the
diameters is required to avoid crystallisation. Section 4 discusses
a thermodynamic property, the pressure. The last
but one section
contains our results for the dynamics, and the last section is a discussion.
All results are for fluids at equilibrium not glasses.
The focus is on the relaxation of the structure
of the fluids, diffusion coefficients and other properties
of motion at times large with respect to the relaxation time
are not calculated.

\section{Simulation methodology}

Conventional NpT Monte Carlo and NVE molecular dynamics techniques are used
\cite{allen87}. $N$, $p$, $V$, $E$, and $T$ are the number of particles,
the pressure, the volume, the energy and the temperature, respectively.
The only nonstandard part concerns the polydispersity of the system,
the presence of spheres of different diameters.
In theoretical treatments of polydispersity first the thermodynamic
limit $N\rightarrow\infty$, is taken and then the number of components
present is taken to $\infty$. This gives rise to
a continuous distribution of sizes of spheres
\cite{salacuse82,gualtieri82} with a number
density $\rho x(\sigma){\rm d}\sigma$ of spheres
in the range $\sigma$ to $\sigma+{\rm d}\sigma$.
$\rho=N/V$ is the total number density of spheres.
The width of the distribution of diameters can
be characterised by a (dimensionless) width parameter $w$.
The larger $w$ is the broader the distribution of sizes present in
the mixture. In the limit $w\rightarrow0$ we recover a monodisperse
system. A simple functional form for the distribution function $x$ is
the hat function:
\begin{equation}
x(\sigma)=\left\{
\begin{array}{ll}
0 & ~~~~~~ \sigma< {\overline\sigma}(1-w/2) \\
({\overline\sigma} w)^{-1} &
 ~~~~~~ {\overline\sigma}(1-w/2)\le \sigma\le{\overline\sigma}(1+w/2)\\
0 & ~~~~~~ \sigma > {\overline\sigma}(1+w/2) \\
\end{array}\right. ,
\label{xs}
\end{equation}
where ${\overline\sigma}$ is the mean diameter.
The standard deviation divided by its mean, of the distribution $x(\sigma)$,
$s=w/\sqrt{12}$.

Simulations are of course of a finite system --- we cannot take the
thermodynamic limit in a simulation.
For finite $N$, $x(\sigma)$ cannot be a continuous function. For
our simulations we generate a set of $N$ diameters by generating
a set of $N$ pseudo-random numbers
in the range ${\overline\sigma}(1-w/2)$ to ${\overline\sigma}(1+w/2)$.
Ideally we would average over many different realisations
of the polydispersity by
performing computer simulations for many different sets of the $N$
diameters. However, as our systems have long relaxation times and so
require long runs this is not feasible. Thus all the results we will present
for a given polydispersity width
will be for a single system of 1372 particles with one set of the 1372
diameters. We have performed some simulations
with different sets of the diameters
but the same number of particles and the results did not
change significantly. We have also performed simulations with 256 and
500 particles and obtained results which did not differ
significantly with the exception of some results for the dynamics.
We discuss these differences below.

NpT Monte Carlo simulations are used to start with and to compress the fluid
phase up to the required density. NVE molecular dynamics are used
to equilibrate at a particular density and to obtain
all the averages shown, including the pressure.
All simulation results are for fluids at (possibly
metastable) equilibrium, not
for glasses. 
Our results are in reduced units. We use the mean diameter
${\overline \sigma}$ as our length scale. The reduced pressure
$p_r$ is then $p_r=Z\rho{\overline\sigma}^3$, where $\rho$ is the number
density of particles and $Z=pV/(NkT)$ is the compressibility
factor. The results are all averages over 8 runs.
We use the variation between the results of the individual runs
to estimate statistical errors. In no case are they significantly
larger than the plot symbols used in the figures.

A time scale can obtained by the ratio of a distance
to a velocity. The velocity we choose is the root mean square
velocity along one of the 3 Cartesian axes,
of a particle with a diameter equal to the mean diameter.
This velocity is equal to $(kT/{\overline m})^{1/2}$, where
${\overline m}$ is the mass of particle with diameter ${\overline\sigma}$.
The time scale is thus ${\overline\sigma}({\overline m}/kT)^{1/2}$
and all times will be made dimensionless by dividing by this.
For molecular dynamics
we have to specify the masses of the particles.
We assume that the mass of a sphere scales with the cube of its
radius, correct if all
spheres are made from the same substance with the
same uniform density.
The density we use is the volume fraction
$\eta=(\pi/6)\rho{\overline{\sigma^3}}$, where ${\overline{\sigma^3}}$
is the third moment of $x(\sigma)$.

\section{Crystallisation}

Although we will present results only for systems which showed no
sign of crystallising, some of our systems did start to crystallise.
Speedy \cite{speedy97} has undertaken a study of the
freezing of monodisperse hard spheres within molecular dynamics simulations.
He finds crystallisation only above a volume fraction of 0.54.
We simulated 1372 monodisperse hard spheres. They were stable for a time
much longer than the relaxation time of the fluid at a volume fraction
$\eta=0.54$ but crystallised at $\eta=0.56$. Thus we agree with the
findings of Speedy.
Simulations of 1372 spheres with
a polydispersity width of $w=0.1$ also crystallised
but when the width was increased to $w=0.2$ ($s=5.8\%$) the fluid
was stable, it never crystallised.
Simulations at polydispersity widths $w=0.3$, 0.5 and 0.7,
of both 256 and 1372 spheres,
never showed any sign of crystallisation. They remained amorphous
up to the highest densities simulated. Widths $w=0.3$, 0.5 and 0.7
correspond to values of $s$ of 8.7\%, 14.4\% and 20.2\%,
respectively.

%He finds that, surprisingly, the density at which an initially
%fluid configuration of the spheres freezes decreases as the system
%size decreases.
%This is over the range $N=250$ to 4000.
%It is surprising as we might expect
%that the probability of a crystalline nucleus forming in a system is
%proportional to its volume.
%It is possible that the periodic boundary conditions
%are encouraging the nucleation of the crystalline phase.
%This uncertainty in how the nucleation rate varies with system
%size makes predicting the lifetime of a bulk sample
%from our simulations impossible. However,

Experiments on colloids find no crystallisation for
polydispersities beyond a standard deviation $s$ close to 8\%
\cite{pusey91,megen98}.
Thus our finding of crystallisation when the standard deviation
is 3\% ($w=0.1)$ but not when it is 6\% ($w=0.2$) is
consistent with experiment. However, our simulations
are for systems much much smaller than experiment and
the time elapsed until nucleation occurs will depend on
system size.
Theoretical and computer simulation studies of the equilibrium
phase behaviour of polydisperse hard spheres
\cite{barrat86,pusey87,mcrae88,bolhuis96,bart99,sear98,bart99b,alm99}
show that polydispersity destabilises the crystalline phase, pushing
the fluid-crystal transition to higher volume fractions.

Our systems with polydispersity widths
of $w=0.2$, 0.3, 0.5 and 0.7
do not crystallise even at volume fractions up to 0.58
because either the fluid phase is still the equilibrium phase even at these
high densities or it is metastable with respect to fluid-crystal
coexistence but the fluid is not
sufficiently deep into the coexistence region for the barrier
to nucleation of the crystalline phase to be small \cite{debenedetti}.
The nucleation rate varies as the inverse of the product of the relaxation
time and the exponential of the free energy barrier \cite{debenedetti}.
If the barrier is large up to densities at which the relaxation time
is very large, say $\eta=0.57$-0.58,
then the rate will always be very small. Perhaps
too small to be observed.

Theoretical studies
\cite{barrat86,pusey87,mcrae88,bart99,sear98} suggest that
for polydispersity widths of around 0.3 and greater,
the crystalline phase cannot form with the full width of the
distribution. It can only form if it accepts only a narrow range of
diameters. Neither in experiments \cite{pusey91,megen98} nor here
is the formation of a crystalline phase with a narrow range
of diameters observed to form from a polydisperse mixture of
spheres. It is not clear whether or not the fact that the crystalline
phase can only form from a fraction of the distribution of diameters
is a cause of the fact that crystallisation is not observed.

We checked for segregation of the large and small spheres, i.e.,
for a tendency to phase separate into a phase of large spheres
coexisting with a phase of small spheres. There was no sign of such
segregation. Segregation is not expected at the polydispersity
widths studied here in the fluid phase. The
Boublik-Mansoori-Carnahan-Starling-Leland (BMCSL)
theory \cite{boublik70,mansoori71}
as generalised to polydisperse hard spheres by Salacuse and
Stell \cite{salacuse82} does
however predict phase separation
at much broader distributions of diameters \cite{cuesta,warren}.
This fluid-fluid separation is metastable with respect to
phase separation plus
crystallisation of the large spheres \cite{sear}.

\section{Pressure}

\begin{figure}
\caption{
%\lineskip 8pt
%\lineskiplimit 8pt
The reduced pressure of polydisperse hard spheres.
The solid, dashed and dotted curves are the predictions of the BMCSL theory
for polydispersity widths of $w=0.3$, $0.5$ and $0.7$, respectively.
The circles, squares and triangles are the results of simulation, for
polydispersity widths of $w=0.3$, $0.5$ and $0.7$, respectively.
}
\label{figpress}
\begin{center}
\epsfig{file=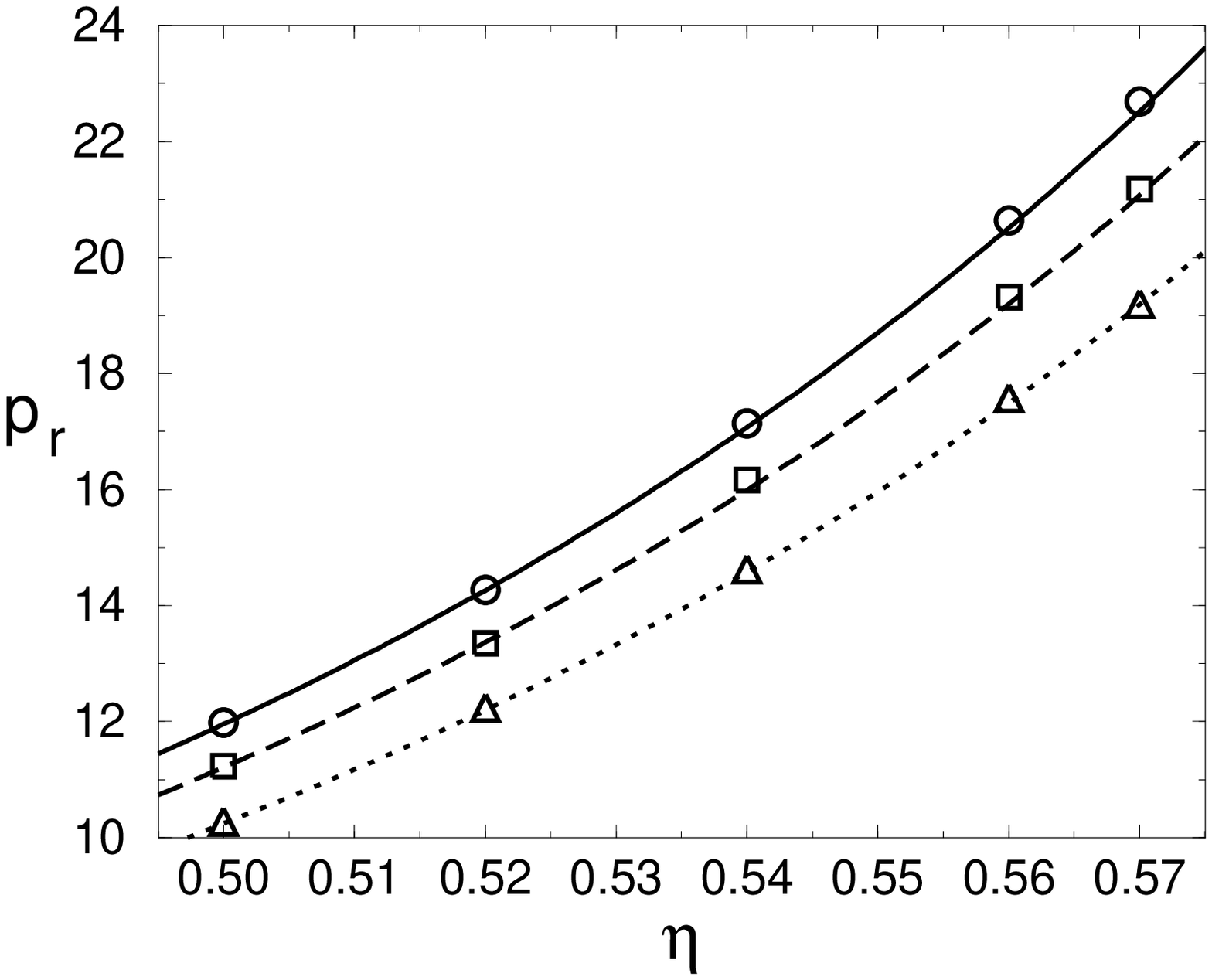,width=3.6in}
\end{center}
\end{figure}

The reduced pressure is plotted as a function of volume fraction in
Fig. \ref{figpress}. Results from simulation are shown along with the
predictions of the BMCSL expression for the pressure
\cite{boublik70,mansoori71,salacuse82}.
The agreement between the BMCSL equation
and simulation is excellent. The only noticeable deviation is that
the BMCSL is a little too high at the highest densities for
$w=0.3$ and 0.5.

The pressure tends to
decrease as the polydispersity width increases.
The random-close-packing density of polydisperse hard spheres
increases as the polydispersity width increases \cite{schaertl94,he99}.
The pressure diverges at random-close-packing. Therefore,
at constant $\eta$ we move farther
from the density where the pressure diverges as the polydispersity width
increases and so it is unsurprising that the pressure decreases.
The distribution used in Refs. \citen{schaertl94,he99} was
different to the hat function used here but the trend should
be relatively insensitive to the exact form of the distribution
when the polydispersity is not too great.

\section{Dynamics}

\begin{figure}
\caption{
%\lineskip 8pt
%\lineskiplimit 8pt
The reduced relaxation time $\tau$ as a function of volume fraction
$\eta$.
The open circles, squares and triangles are the results of simulation, for
polydispersity widths of $w=0.3$, $0.5$ and $0.7$, respectively.
The filled triangles are $\tau^{(s)}$ and
$\tau^{(b}$, the relaxation times
of the the smallest and largest spheres, at
a width of $w=0.7$. The smallest spheres have the smaller $\tau$ at all
densities.
}
\label{figtau}
\begin{center}
\epsfig{file=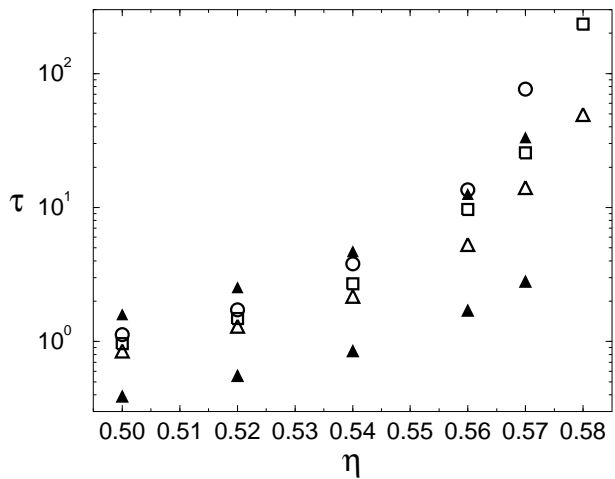,width=3.6in}
\end{center}
\end{figure}

\begin{figure}
\caption{
%\lineskip 8pt
%\lineskiplimit 8pt
The reduced relaxation time $\tau$ as a function of
the reduced pressure $Z$.
The circles, squares and triangles are the results of simulation, for
polydispersity widths of $w=0.3$, $0.5$ and $0.7$, respectively.
The +'s are for monodisperse spheres. For $w=0.3$ and 0.7,
and for monodisperse spheres we have extended the calculations down
to $\eta=0.3$.
}
\label{figtauz}
\begin{center}
\epsfig{file=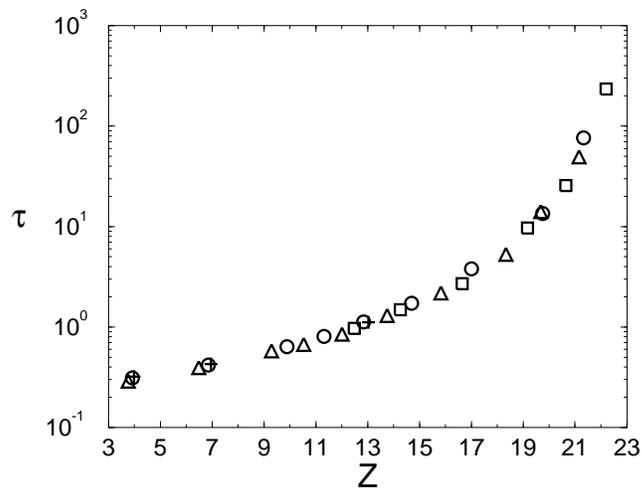,width=3.6in}
\end{center}
\end{figure}

As we are considering dense states the motion of a
particle is strongly restricted by its neighbours. Over some
timescale a particle will rattle back and fore in a cage formed
by its neighbours; then as we are considering a fluid not a glass, the
particle will `break out' of the cage and eventually its motion
will be diffusive.
We can assess whether or not particles are just
rattling back and fore or are moving significant distances, i.e.,
distances comparable to their own diameter, by examining the
intermediate scattering function, sometimes called the self
incoherent intermediate scattering function, $F_s(q,t)$.
As indicated it is a function of wavevector $q$ and time $t$,
it is also a function of volume fraction and
polydispersity. It is defined by \cite{hansen86}
\begin{equation}
F_s(q,t)=\frac{1}{N}\sum_{i=1}^N
\cos({\bf q}.[{\bf r}_i(t)-{\bf r}_i(0)]),
\label{fsq}
\end{equation}
where ${\bf r}_i(t)$ is the position vector for the $i$th particle
at time $t$. For the fluids we consider, it is not a function
of the initial time, taken to be $t=0$ above, only of the time elapsed, $t$.
For our isotropic systems it is also
only a function of the magnitude $q$ of the wavevector
${\bf q}$, not of its orientation. To improve statistics
we average over wavevectors along the $x$, $y$ and $z$ axes.
Clearly $F_s(q,0)=1$ and at times $t$ sufficiently long that the
positions of the particles are no longer correlated with their
positions at time $t=0$, then $F_s(q,t)=0$. As the particles
move, their positions become decorrelated with their positions
at $t=0$ and $F_s$ decays to 0. For a fixed $q$, $F_s(q,t)$ will
become small when the particles have moved a distance of about
$\pi/(2q)$. All our results are for
$q=2\pi/{\overline \sigma}$, and so $F_s$ will become
small when most of the particles have moved a distance of around
a quarter of
their diameter. $F_s$ enables us to define a relaxation
or correlation time for the dynamics, $\tau$. We define $\tau$ by
\begin{equation}
F_s(2\pi/{\overline\sigma},\tau)=1/{\rm e}.
\label{taudef}
\end{equation}
Values of $\tau$ for $w=0.3$, 0.5 and 0.7 are shown in
Fig. \ref{figtau}. 
As expected $\tau$ increases rapidly as the
volume fraction increases; by about 2 orders of magnitude
between $\eta=0.5$ and 0.58. It is also clear that at constant
volume fraction, increasing the polydispersity decreases the
relaxation time $\tau$. However,
if the relaxation time is plotted as a function of
the compressibility factor $Z$,
Fig. \ref{figtauz}, then the data for the 3 polydispersity
widths essentially follow the same curve.
Note that in Fig. \ref{figtauz}
we have included data for monodisperse spheres (at
densities below where they crystallise) and shown results down to $\eta=0.3$.
Spheres with a polydispersity width of $w=0.7$ at $\eta=0.57$
have almost the same $Z$ and almost the same relaxation
time as spheres with $w=0.3$ at $\eta=0.56$. These are the
two points which are almost superposed in Fig. \ref{figtauz}.
To a good approximation the relaxation time as defined
by Eq. (\ref{taudef}) is a function only of the compressibility
factor $Z=pV/(NkT)$.
It is not obvious to the author why $\tau$ should depend only
on $Z$ although there are theories which attempt
to relate dynamic quantities such as $\tau$
to thermodynamic quantities \cite{ediger96,debenedetti}.
If we plot $\tau$
as a function of the reduced pressure $p_r$ (not shown) then the results
for the 3 polydispersity widths do not fall on the same curve,
in particular the results for $w=0.7$ are above those for the
narrower distributions.

We should note that for $\eta=0.56$ and above the relaxation times
calculated using 256 spheres are consistently
above those calculated with 1372 spheres.
For $w=0.3$ and $\eta=0.57$ the relaxation times
are 77 and 110 for 1372 and 256 spheres, respectively.
This suggests that the result for 1372 spheres is
10-30\% higher than that of an infinite system.
When $\tau$ is large,
we observe finite size effects for the dynamics but not for the
statics. This seems to be a quite general finding \cite{kim00}.
It suggests that there is some length scale associated with the
dynamics but not the statics which grows as the relaxation time
$\tau$ increases. Recently this possibility has been extensively
studied by Glotzer and coworkers
\cite{donati99,bennemann99,donati99b,glotzer99}.
They define a purely dynamical correlation length and observe
that it increases
as the dynamics slow down. Perhaps the easiest way to see the correlations
in the dynamics is to plot out the fastest and slowest moving
particles. We simulate the particles with $w=0.3$ at the highest
volume fraction studied for this polydispersity, $\eta=0.57$.
Fig. \ref{figfast} shows the fastest 5\% of the 1372 particles, while
Fig. \ref{figslow} shows the slowest 5\%. Fastest and slowest
are defined as having the largest or smallest displacement
over a time interval of length 29.4. This is a little less than
half the relaxation time $\tau$.
Clearly, neither the
fastest nor the slowest particles are randomly distributed,
both show strong clustering --- the dynamics are highly correlated.
This has been observed in previous simulation studies
\cite{hurley95,kim00,kob97,donati99,bennemann99,donati99b,glotzer99}.

So over times of order $\tau$
the positions of the fastest particles are correlated. The movement
of the particles is very far from being independent diffusion of
each particle. If the particles were diffusing independently
then for each particle, the probability of finding
the particle at a point would be a Gaussian function of the
distance between the point and the position of the particle at $t=0$.
As they are not diffusing independently it will be not be a Gaussian.
We can measure deviations of the distribution of particles from
a Gaussian by using the non-Gaussian parameter introduced
by Rahman \cite{rahman64} and defined by
\begin{equation}
\alpha(t)=\frac{3N^{-1}\sum_{i=1}^N[{\bf r}_i(t)-{\bf r}_i(0)]^4}
{5\left(N^{-1}\sum_{i=1}^N[{\bf r}_i(t)-{\bf r}_i(0)]^2\right)^2}-1.
\label{alpha}
\end{equation}
For a Gaussian distribution the ratio of the fourth moment
to the square of the second is 5/3 and
$\alpha=0$.
In Fig. \ref{figalph} we have plotted
$\alpha$ as a function of time for a number of volume fractions.
For monodisperse particles the Maxwell-Boltzmann
distribution of velocities enforces $\alpha=0$ in the
short time, ballistic, regime, and at times much longer than
$\tau$ we have diffusion and again $\alpha=0$.
For polydisperse particles $\alpha$ is nonzero in the ballistic
regime due to different particles having different masses, and
nonzero at very long times due to larger particles having
smaller diffusion coefficients than smaller particles.
So, in Fig. \ref{figalph} $\alpha$ is always nonzero. The
value of $\alpha$ in the ballistic limit depends only on the
polydispersity width $w$, its value in the other limit
comes from the spread in diffusion coefficients with particle
size and so will also depend on density.
Positive values of $\alpha$ come from a large fourth moment, due
to more particles with large displacements than would be found for
a Gaussian distribution with the same second moment.
It is clear if we compare
Fig. \ref{figalph} with the relaxation times of Fig. \ref{figtau}
that the non-Gaussian parameter $\alpha$ goes through a maximum
at a time comparable to but (at least at the highest densities) less
than the relaxation time $\tau$. This maximum value increases rapidly as
$\tau$ increases. The deviations from a Gaussian are much larger at
a volume fraction of 0.57 than at 0.5.

Recently, experiments have imaged colloidal suspensions of
hard-sphere-like particles at the high volume fractions considered
here \cite{kasper99,kegel00,weeks00}. The suspensions are of particles
with somewhat smaller polydispersities than considered here,
8\% \cite{kasper99},
6\% \cite{kegel00} and 5\% \cite{weeks00}.
The dynamics of colloidal particles \cite{russel},
are rather different from those of free particles
such as our hard spheres. Colloidal particles are immersed
in fluid and so their motion is diffusive on
all time scales on which the particles move a significant distance.
In addition, the motion of nearby particles is coupled
by hydrodynamic interactions through the fluid.
If we compare the $\alpha$ values measured by
Weeks {\it et al.} \cite{weeks00} with those of Fig. \ref{figalph}
we find that the simulation and experimental results are rather
different. The experimental values of $\alpha$ are consistently
larger than those in Fig. \ref{figalph}, Weeks {\it et al.} measure
a maximum $\alpha$ at $\eta=0.52$ of about 1.5 and at $\eta=0.56$
of a little under 2.5. The dynamics on a time scale of
the relaxation time $\tau$ must be rather different for colloidal
particles and free particles, although in both cases the
relaxation time increases very rapidly and in both cases
glasses can form.
%Our particles are more polydisperse than
%in experiment and polydispersity increases $\alpha$ so the
%comparison with experiment may understate the differences
%between the two types of dynamics, although the distribution of the
%diameters in experiment is, of course, not the hat-function used here.
The results of Kegel and van Blaaderen
\cite{kegel00} for $\alpha$ are also larger than our simulation
results; those of Kasper {\it et al.} \cite{kasper99} are much larger,
at $\eta=0.56$ the maximum value of $\alpha$ is over 5.

Two obvious sources of the difference
between molecular dynamics simulation and experiment are:
i) the motion of colloidal particles is diffusive
even on time scales much less than $\tau$, and ii)
hydrodynamic interactions between particles.
i) and ii) could be distinguished by performing
both molecular dynamics and Brownian dynamics
of the same potential \cite{lowen91,gleim98}.
Brownian dynamics is
an approximation to the dynamics of particles in a fluid which
accounts for the fact that particles diffuse even over very small
length scales but neglects hydrodynamic interactions between particles
\cite{russel}. 

Another way to look at the large spread in the amount individual particles
are moving is to define $F_s$'s for each particle, and then
use this to obtain a relaxation time for each particle.
Averaging over wavevectors along the $x$, $y$ and $z$
axes we have for the $i$th particle
\begin{eqnarray}
\frac{1}{3}\left\{
\cos(q[x_i(\zeta_i)-x_i(0)])+ \cos(q[y_i(\zeta_i)-y_i(0)])+\right.\nonumber\\
\left. \cos(q[z_i(\zeta_i)-z_i(0)]) \right\}
=1/{\rm e},
\end{eqnarray}
which defines the relaxation time for the $i$th particle, $\zeta_i$.
As always $q=2\pi/{\overline\sigma}$.
In Fig. \ref{figzeta} we have plotted the
probability function $P(\zeta)$, where $P(\zeta){\rm d}\zeta$
is the probability that a particle will have a relaxation
time between $\zeta$ and $\zeta+{\rm d}\zeta$.
For both $\eta=0.5$ and 0.56 there is a peak
at short times and then $P$ decays. The decay is clearly close
to exponential, which makes sense as (at not-too-short times)
the probability of a particle leaving its cage in some short
time interval should be independent of time; at equilibrium the
environment of a particle or cluster of particles is on average
not changing with time.
For $\eta=0.5$ the
peak is a little below $\tau$ at that density but for $\eta=0.56$
the peak is at about one tenth of $\tau$. At the higher density the
probability distribution of $\zeta$ is very broad.

\subsection{Polydispersity}

\begin{figure}
\caption{
%\lineskip 8pt
%\lineskiplimit 8pt
The slowest 5\% of our particles, 69 out 1372, are drawn
as spheres of diameter ${\overline\sigma}$.
Slowest is defined as having traveled the smallest
distance within a time interval of 29.4.
The volume fraction $\eta=0.57$ and $w=0.3$.
}
\label{figslow}
\begin{center}
\epsfig{file=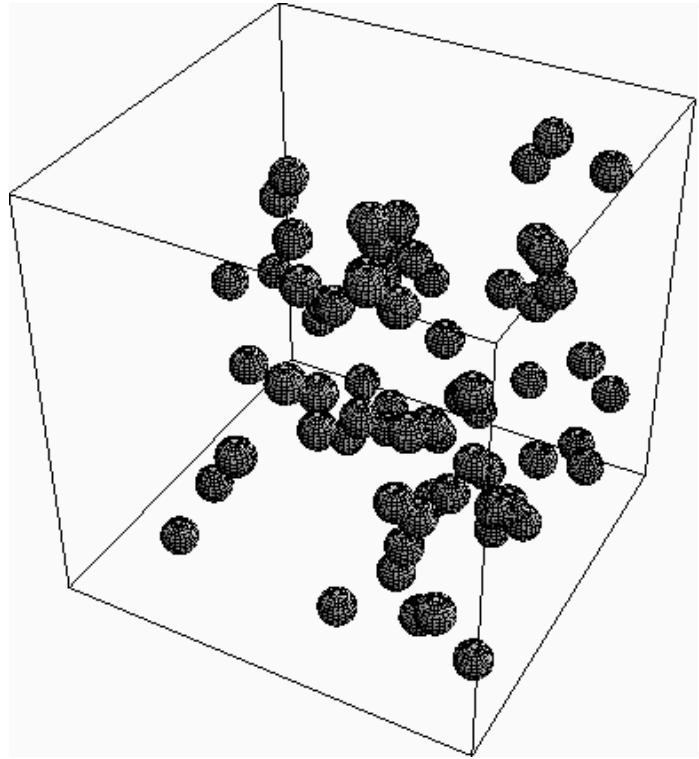,width=3.6in}
\end{center}
\end{figure}

The spheres are polydisperse, different spheres
have different diameters and so their static and
dynamic properties will differ. We expect the smaller spheres
to move larger distances than larger spheres in the same time.
At very short times, in the ballistic regime, this is trivially
true due to the smaller mass and hence higher velocity of the smaller
spheres. We can look at motion over a time scale of order $\tau$
by examining the sizes of the fastest and slowest 5\%
of the spheres. Figs. \ref{figfast} and \ref{figslow} show
these spheres for $w=0.3$. We average over 8 runs with an
average time interval close to that in Figs. \ref{figfast} and \ref{figslow}.
The average diameter of the fastest 5\% (69) of the spheres,
over a time interval of 29.2, is
$0.95{\overline\sigma}\pm0.01{\overline\sigma}$,
and for the slowest 5\% it is $1.01{\overline\sigma}\pm0.01{\overline\sigma}$.
The effect is small but as we would expect the fastest spheres are smaller
than average. For a polydispersity width of 0.7, over a time interval
of 5.88, the average diameter of the fastest 5\% is
$0.79{\overline\sigma}\pm0.02{\overline\sigma}$,
and for the slowest 5\% it is $1.08\pm0.02{\overline\sigma}$.
A time of 5.88 is, as with the less polydisperse spheres, somewhat
less than half $\tau$ and so near the maximum in $\alpha$.
Of course the effect is now larger but still the slowest spheres
are not much larger than the mean, their mean diameter would
have been above 1.3 if all the slowest spheres were also the largest.
The difference between the mean diameter of the fastest spheres
and the overall mean diameter is twice the difference
between the mean diameter of the slowest spheres and the overall
mean diameter.

We have defined an
intermediate scattering function, $F_s$, for all spheres
regardless of diameter in Eq. (\ref{fsq}). We can also define
an intermediate scattering function
for subsets of the particles with diameters in some range
$\sigma_{min}$ to $\sigma_{max}$, $F_s(q,t;\sigma_{min},\sigma_{max})$.
The definition is completely analogous to that of
the total $F_s(q,t)$ in Eq. (\ref{fsq}),
\begin{eqnarray}
F_s(q,t;\sigma_{min},\sigma_{max})=~~~~~~~~~~~~~~~~~~~~~~~~~~~~~~\nonumber\\
\frac{1}{N_{mm}}
\sum_{i=1} ^{N_{mm}}
\cos({\bf q}.[{\bf r}_i(t)-{\bf r}_i(0)]),\nonumber\\
\label{fsqmm}
\end{eqnarray}
where the sum is over all spheres with diameters in the
range $\sigma_{min}$ to $\sigma_{max}$, and $N_{mm}$
is the number of spheres in this range.
As before $q=2\pi/{\overline\sigma}$ and we average
over wavevectors along the $x$, $y$ and $z$ axes.
We will study intermediate scattering functions for the smallest,
$F_s^{(s)}(q,t)$, and largest spheres, $F_s^{(l)}(q,t)$, defined by
\begin{eqnarray}
F_s^{(s)}(q,t)&=&
F_s(q,t;{\overline\sigma}(1-w/2),
{\overline\sigma}(1-w/2+0.05))
\nonumber\\
F_s^{(l)}(q,t)&=&
F_s(q,t;{\overline\sigma}(1+w/2-0.05),{\overline\sigma}(1+w/2)).
\nonumber\\
\end{eqnarray}
These are for the particles
within $0.05{\overline\sigma}$ of the minimum
or maximum diameter. For
$w=0.7$ this corresponds to spheres with diameters
in the ranges 0.65 to $0.7{\overline \sigma}$ and
1.3 to $1.35{\overline\sigma}$.
We also define relaxation times $\tau^{(s)}$ and $\tau^{(l)}$
from $F_s^{(s)}(q,t)$ and $F_s^{(l)}(q,t)$, respectively, using
the analogues of Eq. (\ref{taudef}). $\tau^{(s)}$ and $\tau^{(l)}$
for $w=0.7$ are plotted as filled triangles in Fig. \ref{figtau}.
The ratio $\tau^{(l)}/\tau^{(s)}$ increases with increasing density but
slower than $\tau$ does. At $\eta=0.57$ the ratio is approximately
10.

A feature of the dynamics when $\tau$ is large is that they are
heterogeneous, some particles travel much farther than others
in a time $\tau$. In a fluid of monodisperse particles
these heterogeneities are dynamic, a particle may be fast at one time
but later on may be slow. For polydisperse particles in
addition to this source of heterogeneity there is the spread
in diameters which means that some particles, the smaller ones,
are on average faster than others over all time scales.
In Fig. \ref{figalph} we can compare the
$\alpha$ parameter for $w=0.3$ and 0.7 at the same volume fraction
$\eta=0.57$. $\alpha$ is larger for the more polydisperse
spheres; this is the case despite the fact that the
relaxation time for the more polydisperse particles is smaller.
The system with $w=0.7$ and $\eta=0.57$ is at almost the
same compressibility factor
as the one with $w=0.3$ and $\eta=0.56$ so we see that
at constant compressibility factor
$\alpha$ increases sharply with polydispersity.

For weakly polydisperse spheres the relaxation times for
all sizes of sphere increase
together and so if we are at high enough density then
$\tau$ will be very large and on a timescale much less than $\tau$
all spheres will be localised, none or at least very few
will have moved more than a small fraction of their diameter.
However, if spheres with a wide range of diameters are
present then the smallest particles will have a relaxation time 
scale which is much less than
$\tau$. Then over some time scale much less than $\tau$
but much larger than the relaxation time
scale for the smallest particles not all the spheres will
be localised. The smallest spheres will be diffusing.
This has been observed in binary mixtures of hard spheres in
which the smaller spheres are much smaller than the larger spheres,
by Jackson {\it et al.} \cite{jackson87}. The same effect is
seen in crystals of binary mixtures \cite{ermak81}. This is
an extreme example of polydispersity making the dynamics
even more heterogeneous than they are when the particles are
all the same size. Note that polydispersity and the heterogeneous
nature of the dynamics when $\tau$ is large have the same effect,
they produce a wide spread in speeds/relaxation times of the
individual particles.
Even for monodisperse particles, for times much
less than $\tau$ the heterogeneous dynamics mean that some particles
have left their cages. However, if we have some particles much smaller than
others then the small ones may have a
relaxation time not much larger
and a diffusion constant \cite{jackson87} not much smaller than
at lower densities.

\section{Discussion}

We have studied the dynamics of dense polydisperse hard spheres
using molecular dynamics. The relaxation time $\tau$
at a given volume fraction was found to depend
on the polydispersity, it decreased as the polydispersity increased.
So, the glass transition is pushed to higher volume fractions
as the spheres become more polydisperse. This is consistent
with the increase in the density of random-close-packing
with increasing polydispersity \cite{schaertl94,he99}.
Although we have found this using molecular dynamics, we expect
that experiments on polydisperse colloids would show that as
polydispersity increased the kinetic glass transition observed
in experiment would move to higher volume fractions.

\begin{figure}
\caption{
%\lineskip 8pt
%\lineskiplimit 8pt
The non-Gaussian parameter $\alpha$ as a function of time.
The circles and squares are for a polydispersity widths $w=0.3$ and $0.7$,
respectively. The three curves for $w=0.3$ are from bottom to
top, for $\eta=0.52$, 0.56 and 0.57. The squares are for $\eta=0.57$.
}
\label{figalph}
\begin{center}
\epsfig{file=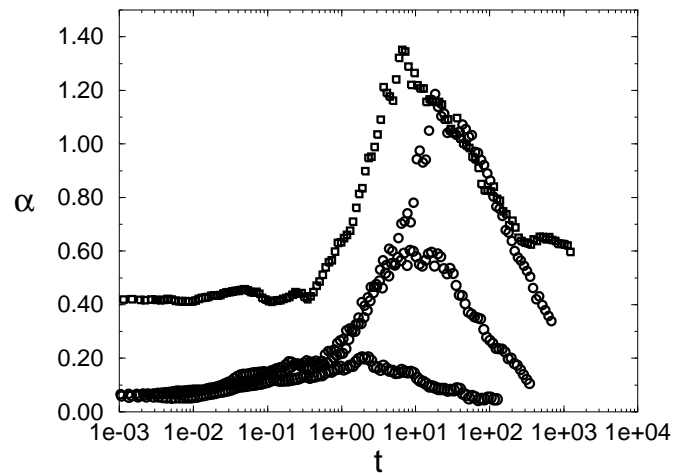,width=3.6in}
\end{center}
\end{figure}

The relaxation time $\tau$ is, to a good approximation, a function only of
the compressibility factor $Z$, changing the polydispersity at fixed $Z$
has no effect on $\tau$. This is despite the fact that
a characteristic of the dynamics, such as $\alpha$, changes
a great deal at fixed $Z$ as $w$ increases; compare the
$w=0.3$, $\eta=0.56$ and $w=0.7$, $\eta=0.57$ curves in Fig. \ref{figalph},
which have almost the same $Z$ but very different $\alpha$'s.
Apparently, increasing the polydispersity
at constant volume fraction reduces the relaxation time by
reducing $Z$. The virial equation \cite{plischke},
\begin{equation}
Z=1-\frac{1}{6NkT}\langle\sum_{i\ne j}{\bf r}_{ij}.{\bf f}_{ij}\rangle
\label{virial}
\end{equation}
relates $Z$ to the forces between the particles.
$\langle\rangle$ denotes an ensemble average.
In Eq. (\ref{virial})
${\bf r}_{ij}$ and ${\bf f}_{ij}$ are the vector between the centres
of the $i$th and $j$th particles and the force
on the $i$th particle due to the $j$th particle, respectively.
The sum is over all pairs of
particles. So, $Z$ is a constant, 1, plus a term
proportional to the sum over
the product of the interparticle forces and the interparticle
separations. Why the relaxation time should be a function only
of this product is not clear to the author.

Although $Z$ can be expressed in terms of forces it is also of course
a thermodynamic quantity. At constant temperature $Z$ varies as
$p/\rho$ so $\tau$ being a function of $Z$ is equivalent
to it being a function of the ratio of the pressure to the
number density.
As the energy of hard spheres is purely
kinetic it is independent of volume, thus the pressure
is simply the volume derivative of the entropy times the temperature.
So, the relaxation time is a function only of $Z$ and so
is a function only of the ratio of
volume derivative of the entropy to the number density.
But it is not a simple function
of it; $\tau$ does not vary as the exponential
of the $Z$, it increases more rapidly.
The pressure shows no sign of a discontinuity in slope so
we conclude that there is no phase transition in the density
range studied here. It is an open question whether or not
there is a phase transition to an `ideal' glass phase at
higher densities, as assumed by Speedy \cite{speedy98},
see also Refs. \citen{ediger96,debenedetti}.

\begin{figure}
\caption{
%\lineskip 8pt
%\lineskiplimit 8pt
The fastest 5\% of our particles, 69 out 1372, are drawn
as spheres of diameter ${\overline\sigma}$ (all the spheres are drawn
the same size). Fastest is defined as having traveled the largest
distance within a time interval of 29.4.
The volume fraction $\eta=0.57$ and $w=0.3$.
}
\label{figfast}
\begin{center}
\epsfig{file=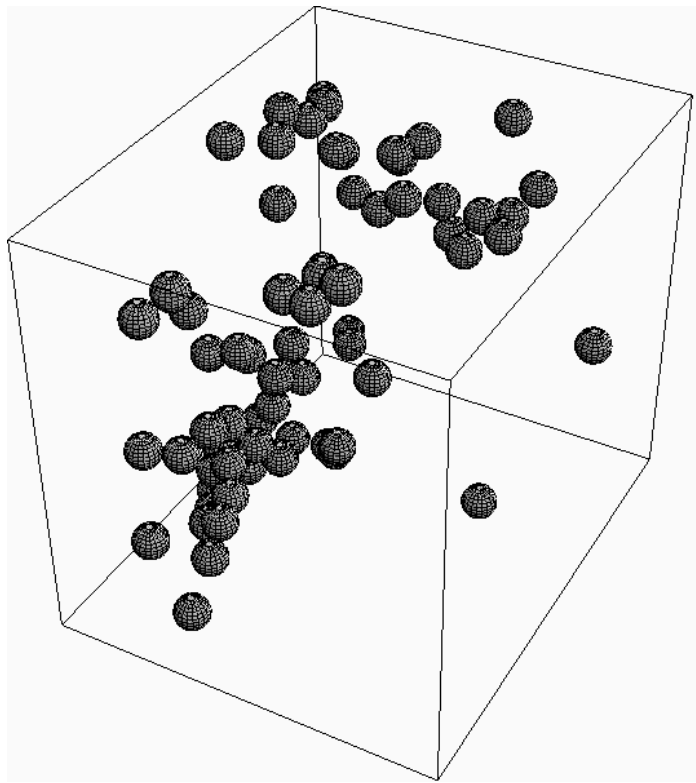,width=3.6in}
\end{center}
\end{figure}

\begin{figure}
\caption{
%\lineskip 8pt
%\lineskiplimit 8pt
The probability distribution function for
relaxation times of individual particles.
The crosses and squares are for $\eta=0.5$ and 0.56, respectively; $w=0.3$.
}
\label{figzeta}
\begin{center}
\epsfig{file=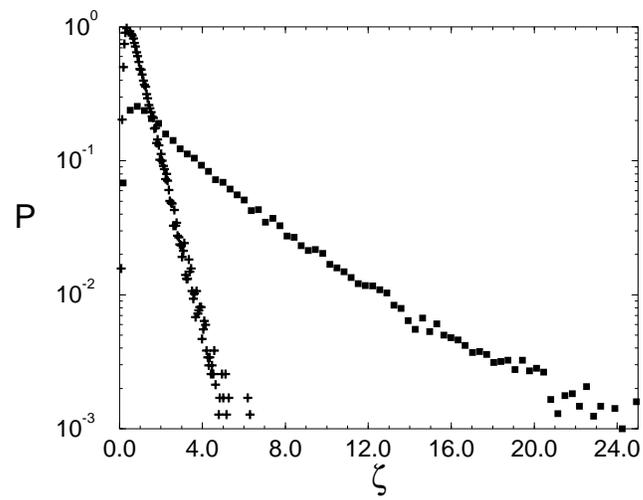,width=3.6in}
\end{center}
\end{figure}

We have characterised the dynamics using the non-Gaussian
parameter $\alpha$, Fig. \ref{figalph}, and the distribution
of relaxation times $P(\zeta)$, Fig. \ref{figzeta}. The
results for $\alpha$ have been compared to experimental
results for colloidal suspensions \cite{kasper99,kegel00,weeks00}.
The dynamics
are heterogeneous \cite{hurley95,donati99b,kob97,glotzer99}
by which we mean that
there is a broad distribution of speeds and of relaxation times
of the particles. This is true not only at times of order
the relaxation time $\tau$ but also, at higher density and
hence $\tau$, for times
at least an order of magnitude larger, see Fig. \ref{figzeta}.
Indeed given the very wide spread of individual relaxation times
$\zeta$ it is clear that the overall relaxation time $\tau$ is
inadequate to characterise the time dependence of the dynamics.
In particular, relaxation is far from complete even at
times much longer than $\tau$.
Comparison of our non-Gaussian parameter with that
obtained in experiments on colloidal hard spheres showed significant
differences. One possible explanation of this is that hydrodynamic
interactions between colloidal particles are acting to make the
relaxation more cooperative, i.e., acting to increase the clustering
shown in Fig. \ref{figfast}. It is known that hydrodynamic
interactions tend to favour collective over relative motion \cite{russel}.

%It is a pleasure to thank J. Cuesta for
%useful discussions.

%I acknowledge the support of EPSRC (GR/N22496).

%\newpage

\end{multicols}

\clearpage

\end{document}